\documentclass[fleqn,twoside]{article}
\usepackage{espcrc2}
\usepackage{amsfonts}
\newif\ifpdf
\ifx\pdfoutput\undefined\pdffalse\else\pdfoutput=1\pdftrue\fi
\ifpdf
  \usepackage[backref,citecolor=blue]{hyperref}
  \let\myhref=\href\def\href#1#2{\penalty-20\myhref{#1}{\tt #2}}%
\else%
  \def\href#1#2{{\penalty-20\tt #2}}
\fi
\def\epspdffile#1{\leavevmode\ifpdf\epsffile{#1.pdf}%
  \else\epsffile{#1.eps}\fi}
\usepackage{epsfig}
\def\asqtad{{\scshape asqtad}}
\def\qcdoc{{\scshape qcdoc}}
\def\M{{\cal M}}			    

\def\defn{\equiv}			    
\def\k{\kappa}				    
\def\mq{m}				    
\def\det{\mathop{\rm det}}		    
\def\tr{\mathop{\rm tr}}		    
\def\Z{{\mathbb Z}}			    
\def\trjlen{\tau}			    
\def\dt{\delta\tau}			    
\def\MD{molecular dynamics}		    
\def\HMC{HMC}				    
\def\RHMC{RHMC}				    
\def\rmsub#1#2{#1_{\mbox{\tiny #2}}}	    
\def\opt#1{\rmsub{#1}{opt}}		    
\def\ninv{\rmsub{N}{inv}}		    
\def\pacc{\rmsub{P}{acc}}		    

\title{Accelerating Fermionic Molecular Dynamics}

\author{M. A. Clark\address[MCSD]{School of Physics, 
        The University of Edinburgh, \\ 
        Edinburgh EH9 3JZ, United Kingdom}
        and
        A. D. Kennedy\addressmark}
\begin{document}

\begin{abstract}
\noindent We consider how to accelerate fermionic molecular dynamics algorithms
by introducing $n$ pseudofermion fields coupled with the $n$th root of the
fermionic kernel. This reduces the maximum pseudofermionic force, and thus
allows a larger molecular dynamics integration step size without hitting an
instability in the integrator.{\parfillskip=0pt\par}
\end{abstract}

\maketitle

\section{Introduction}
For over fifteen years the algorithm of choice for generating lattice field
theory configurations including the dynamical effect of fermions has been
Hybrid Monte Carlo (HMC) \cite{duane87a}. Unfortunately the cost of this
algorithm increases rapidly as the fermion mass \(\mq\) decreases; in order to
keep the \HMC\ acceptance rate \(\pacc\) constant the \MD\ integration step
size \(\dt\) has to be reduced, and for \MD\ trajectories of length
\(\trjlen=1\) this corresponds directly to an increased number of \MD\
integration steps and hence larger cost.

This required decrease in step size is because of the breakdown of symmetric
symplectic integrators. For light dynamical fermions there is an instability
for a few isolated light fermion modes, whose frequency is well separated from
the bulk of the modes.  This instability is seen to be directly responsible for
the exponential decrease of \HMC\ acceptance with integration step size above
some critical value.

We introduce a method for reducing the severity of this problem by reducing the
highest ``effective frequency'' of the fermionic modes, or equivalently of
decreasing the magnitude of the fermionic contribution to the force acting on
the gauge fields. The basic idea follows the suggestion of
Hasenbusch~\cite{Hasenbusch:2001ne} to split the fermionic action into two
parts, and to introduce separate pseudofermion fields for each part. Our
approach can be easily generalised to an arbitrary number of pseudofermion
fields.\footnote{Hasenbusch's method also allows an arbitrary number of
pseudofermion fields to be used, but the parameters in the action have to be
tuned to ensure that the contributions to the force are divided up equally.}

\section{Non-linearity of CG} \label{sec:CG}

We observe that the force due to the fermion kernel \(\M^{-1}\) is dominated by
the smallest eigenvalues of \(\M\). The condition number \(\k(\M)\) is the
ratio of the largest eigenvalue to the smallest eigenvalue, and to a first
approximation controls the rate of convergence of iterative Krylov space
solvers. The largest eigenvalue remains approximately constant as the fermion
mass \(\mq\) is decreased, and the smallest eigenvalue is typically of the
order \(\mq^\alpha\) where \(\alpha\) is \(1\) or~\(2\), so we expect
\(\k(\M)\propto \mq^{-\alpha}\).

Consider the numerical solution of the linear system \(\M\chi=\phi\), where
\(\M>0\) and has condition number \(\k(\M)\).  The cost of solving these linear
equations is proportional to \(m^{-\alpha}\).  On the other hand we could
equivalently solve the set of coupled linear equations \(\sqrt{\M}\chi=\psi\)
and \(\sqrt{\M}\psi=\phi\), each of which has condition number \(\k(\sqrt
M)=\sqrt{\k(\M)}\),\footnote{For \(\M>0\) the positive square root is uniquely
defined.} leading to a cost of order \(2\mq^{-\alpha/2}\) in this case, which
is cheaper for sufficiently small \(\mq\). This reflects the essential
non-linearity of Krylov space solvers.  Indeed, we may even be more adventurous
and solve the set of \(n\) coupled systems \(\root n\of\M\psi_j =\psi_{j+1}\),
where \(\psi_0=\chi\) and \(\psi_n=\phi\), for which we have to perform \(n\)
solves each with condition number \(\k(\M^{1/n}) =\k(\M)^{1/n}\), leading to a
total cost of order \(n\k(\M)^{1/n}\).

Unfortunately we cannot take advantage of this non-linearity, the problem being
that it is not straightforward to apply \(\sqrt\M\) to a vector when \(\M\) is
not serendipitously a manifest square. There are efficient techniques for
evaluating matrix functions, such as computing the optimal polynomial or
rational Chebyshev approximation to the function over the spectrum of the
matrix \cite{kennedy:2003a}. For the rational case the approximations may be
found using the Remez algorithm, and they usually converge exponentially in the
degree of the rational function.  In practice only a relatively low degree
rational function is needed to achieve machine floating-point precision
\cite{Clark:2003na}. Furthermore, if we take a rational approximation then we
can express it as a partial fraction, and apply all the terms simultaneously
(in the same Krylov space) using a multi-shift solver. This reduces the cost of
solving \(\M^{1/n}x=b\) to about the same cost as solving \(\M x=b\). This in
turn is expected to be proportional to the condition number \(\k(\M)\). Sadly,
this means that the cost of the proposed method is of order \(n\k(\M)\) rather
than \(n\k(\M)^{1/n}\), and we are worse off than when we started.

Although this method is clearly useless to accelerate the convergence of Krylov
space solvers, it still does significantly reduce the condition number of each
of the \(n\) solves, and this is what we shall make use of to decrease the
pseudofermionic force. Indeed, we expect the force to be of order
\(n\k(\M)^{1/n}\dt\), which is small compared to \(\k(\M)\dt\) for large
\(\k(\M)\).  A na{\"\i}ve calculation minimising \(n\k(\M)^{1/n}/\k(\M)\) leads
to the conclusion that the optimal number of pseudofermion fields should
be\footnote{Strictly speaking \(\opt{n}\in\Z\), so it must be either \(\lfloor
\ln\k(\M)\rfloor\) or \(\lceil\ln\k(\M)\rceil\) .}  \(\opt{n}= \ln\k(\M)\).

\section{Pseudofermion Sampling}

Recall that we represent the fermion determinant as a pseudofermion Gaussian
functional integral, \(\det\M \propto \int d\phi\,d\phi^\dagger\, \exp{\left(-
\phi^\dagger\M^{-1}\phi\right)}\), and then select a single equilibrium
pseudofermion configuration using a Gaussian heatbath. We expect, therefore,
that the variance of this stochastic estimate of the fermion determinant will
lead to statistical fluctuations in the fermionic force: in other words the
pseudofermionic force may be larger than the exact fermionic force, which is
the functional derivative \(\partial\tr\ln\M(U)/\partial U\) with respect to
the gauge field \(U\). This means that the pseudofermionic force may trigger
the instability in the symplectic integrator even though the exact fermionic
force would not.

An obvious way of ameliorating this effect is to use \(n>1\) pseudofermion
fields (which we shall call \emph{multipseudofermions}) to sample the
functional integral representing the fermion determinant, and this is achieved
simply by writing
\begin{eqnarray*}
  \det\M & = &[\det\M^{1/n}]^n \nonumber \\
  & \propto & \prod_{j=1}^n d\phi_j\, d\phi^\dagger_j\,
    \exp{\left(-\phi^\dagger_j\M^{-1/n}\phi_j \right)};
  \label{eq:multipseudo}
\end{eqnarray*}
that is, introducing \(n\) pseudofermion fields \(\phi_j\) each with kernel
\(\M^{-1/n}\).

We may now follow a similar argument to that in \S\ref{sec:CG} to estimate the
optimal value for \(n\). We must keep the maximum force fixed so as to avoid
the instability in the integrator, so we may increase the integration step size
to \(\dt'\) such that \(n\k(\M)^{1/n}\dt' = \k(\M)\dt\). At constant trajectory
length, and hence constant autocorrelation time, the cost of an \HMC\
trajectory is proportional to the step size, and thus is minimised by choosing
\(n\) so as to minimise \(\dt'/\dt = \k(\M)^{1-1/n}/n\), which leads to the
condition \(\opt{n} \approx \ln\k(\M)\), corresponding to cost reduction by a
factor of \(\dt/\dt' \approx e\ln\k(\M)/\k(\M)\).

\section{\RHMC\ Acceleration} \label{sec:RHMC}

Our method is to apply the \RHMC\ algorithm \cite{Clark:2003na} to generate
gauge field and pseudofermion configurations distributed according to the
probability density \[P(U,\phi_1,\ldots,\phi_n) = \frac1Z \exp{\bigl[-S_B(U) -
S_F(\M)\bigr]},\] where \(S_B\) is the bosonic (pure gauge) action and
\(S_F(\M) = \sum_{j=1}^n \phi^\dagger_j \M^{-1/n}\phi_j\). Optimal rational
approximations are used to evaluate these matrix functions, and we proceed as
we would for conventional \HMC~\cite{duane87a}. Using a multi-shift solver the
computational cost is very similar to \HMC, with the additional overhead of
having to perform a matrix inversion to evaluate the heatbath.

We have performed tests of the algorithm using small lattices with four
flavours of na\"{\i}ve staggered fermions. As well as being the least
computationally demanding simulations to perform, the improvement for such
systems should be a lower bound on the improvement for large volume and/or
Wilson type fermion simulations, where the fermion matrix is less
well-conditioned.

The \(n=1\) case runs were performed using conventional \HMC, and the \(n>1\)
runs using \RHMC. We define an efficiency measure \(E\defn \langle\pacc
\rangle/\ninv\), where \(\ninv\) is number of Krylov solves performed per
trajectory. For \HMC\ this is given by \(1+\trjlen/\dt\) and for \RHMC\ by
\((2+\trjlen/\dt)n\), where \(\trjlen\) is the trajectory length and \(\dt\)
the step size. The additional inversion per field for \RHMC\ is for the
pseudofermion heatbath.

\begin{figure}[t]
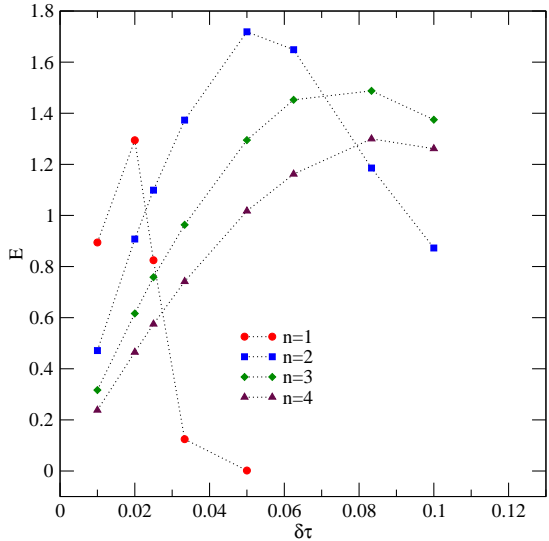

  \epsfxsize=0.45\textwidth
  \centerline{\epspdffile{eff-dt-1}}
  \vskip-6ex
  \caption[Efficiency 1]{Efficiency \(E\) as a function of integration step
    size \(\dt\), for QCD with \(4\) flavours of na\"{\i}ve staggered quarks
    of mass \(\mq=0.025\) at \(\beta=5.26\) on an \(8^4\) lattice with
    trajectory length \(\trjlen=1\).}
  \label{fig:eff-dt-1}
\end{figure}

Figure \ref{fig:eff-dt-1} is a plot of the efficiency against step size for
various value of \(n\). The conjecture that there is some optimal value of
\(n\) is confirmed, and it can be seen that \(n=2\) is the value for this
particular lattice. It represents an increase in efficiency of 33\% (\(n=2\)
peak divided by \(n=1\) peak). We have conducted the same test with a mass
parameter of \(\mq=0.01\), which shows the same optimal value of \(n\), but
with a substantial efficiency increase of \(60\%\). This confirms that the
improvement factor increases as less well-conditioned systems are studied.

\section{Conclusions} \label{sec:conclusions}

We have demonstrated that our method for accelerating the Monte Carlo
acceptance test improves the efficiency over conventional \HMC. This is clearly
still a very preliminary study, and shall be extended to large scale systems
for both {\asqtad} and domain wall simulations using \qcdoc. On such systems,
we would expect the improvement in efficiency to increase, and it will be
interesting to see how much improvement can be gained. It would also be
interesting to compare our method with that proposed by Hasenbusch
in~\cite{Hasenbusch:2001ne}.

\section*{Acknowledgements}

ADK would like to thank Brian Pendleton and Artan Bori\c{c}i for useful
discussions and encouragement.

\end{document}